\documentclass[reprint,amsmath,amssymb,aps,prl,showpacs,floatfix,superscriptaddress]{revtex4-1}

\usepackage{color,graphicx,bm}

\begin{document}

\title{Trion formation in a two-dimensional hole-doped electron gas}

\author{G.\ G.\ Spink}
\affiliation{TCM Group, Cavendish Laboratory, University of Cambridge,
     19 J.\ J.\ Thomson Avenue, Cambridge CB3 0HE, United Kingdom}
\affiliation{Department of Chemical Engineering, University of
     Chester, Thornton Science Park, Chester CH2 4NU, United Kingdom}

\author{P.\ L\'opez R\'\i os}
\affiliation{TCM Group, Cavendish Laboratory, University of Cambridge,
     19 J.\ J.\ Thomson Avenue, Cambridge CB3 0HE, United Kingdom}

\author{N.\ D.\ Drummond}
\affiliation{Department of Physics, Lancaster University, Lancaster
     LA1 4YB, United Kingdom}

\author{R.\ J.\ Needs}
\affiliation{TCM Group, Cavendish Laboratory, University of Cambridge,
     19 J.\ J.\ Thomson Avenue, Cambridge CB3 0HE, United Kingdom}

\date{July 18, 2016}

\begin{abstract}
  The interaction between a single hole and a two-dimensional,
  paramagnetic, homogeneous electron gas is studied using diffusion
  quantum Monte Carlo simulations.  Electron-hole relaxation energies,
  pair-correlation functions, and electron-hole center-of-mass
  momentum densities are reported for a range of electron-hole mass
  ratios and electron densities.  We find numerical evidence of a
  crossover from a collective excitonic state to a trion-dominated
  state in a density range in agreement with that found in recent
  experiments on quantum-well heterostructures.
 \end{abstract}


\maketitle

The rich physics arising from the Coulomb attraction between electrons
and holes in layered semiconductor systems continues to generate
fundamental and technological interest.  Collective many-body effects,
such as the Fermi-edge singularities (FES) in absorption spectra
predicted by Mahan \cite{Mahan}, dominate at high carrier densities,
while excitonic species form in dilute systems \cite{Huard}.  Neutral
excitons, consisting of bound electron-hole pairs, and charged trions,
which are bound states of two electrons and one hole, or two holes and
one electron, are elementary quasiparticles that can be created via
photoexcitation or by chemical or electrical means in a wide range of
materials.  Many optoelectronic devices, from photovoltaics and
light-emitting diodes \cite{LEDs} to optoelectronic storage devices
\cite{Lundstrom}, interconnects and switches \cite{optic,Grosso},
exploit excitonic effects, as does nature in, for example, plant
photosynthesis.

Modulation doped or gated semiconductor quantum wells, including GaAs
wells with AlGaAs barriers and InGaAs/InAlAs junctions, offer
convenient experimental access to high-mobility electron gases.
Control of carrier density and creation of electron-hole pairs via
photoexcitation are readily achieved and early studies identified the
FES \cite{Chemla,Livescu} at high carrier densities.  Kheng \textit{et
al.}\ \cite{Kheng} identified negatively charged trions in CdTe
quantum wells at low carrier densities, 35 years after their
prediction by Lampert \cite{Lampert}.  Particle confinement in a
quasi-two-dimensional well increases the overlap of the hole and electron
wave functions and increases the binding energy of the hole compared
to the bulk semiconductor \cite{Stebe}, making this an ideal system in
which to study trions.  The crossover between high-density FES
dominated by many-body correlations, which we refer to as the
collective excitonic state, and the low-density behavior,
characterized by the presence of excitons and trions, has been
investigated experimentally \cite{Huard,Yusa,Bar-Joseph,Yamaguchi}.

Huard \textit{et al.}\ \cite{Huard} observed a gradual change in the
absorption spectrum of modulation doped CdTe semiconductor quantum
wells from discrete excitonic peaks at low carrier density to broad
FES at high density.  Rapid changes in line shapes and transition
energies seen in the absorption and photoluminescence spectra of a
gated modulation-doped GaAs quantum well allowed Yusa \textit{et al.}\
\cite{Yusa} to locate a critical ``crossover'' density.  Similar
methods were used by Bar-Joseph \cite{Bar-Joseph}.  Yamaguchi
\textit{et al.}\ \cite{Yamaguchi} recently showed that the
photoluminescence linewidth of a gated, undoped GaAs quantum well as a
function of energy shift in a perpendicular electric field can be used
to measure the spatial extent of the trion at a given carrier density.
The measured trion radius increases sharply above a critical density,
which is identified as the crossover.  A theoretical
description of the crossover was given by Hawrylak \cite{Hawrylak}
via an approximate treatment of electron-electron
interactions.

Despite the wealth of experimental information, interpretation of
spectroscopic data is often not straightforward and the properties of
excitonic states are still debated.  Emission of photons at the
anticipated exciton frequencies does not unambiguously signal the
presence of excitons, as a system
formed by an electron gas and a hole will resonate at the exciton
frequency due to many-body interactions \cite{Koch}.  In addition,
experimental samples exhibit great sensitivity to variables such as
temperature, finite quantum-well width, and the presence of disorder
and localization effects from the, albeit spatially removed, dopants
in modulation-doped quantum wells.  Further theoretical insight
into such electron-hole systems is urgently needed.  We have
therefore performed variational and diffusion quantum Monte Carlo (VMC
and DMC) \cite{Review} calculations to understand the important limit
of a zero-temperature, two-dimensional (2D) system comprising a single
hole immersed in a 2D homogeneous electron gas (HEG) interacting via
the Coulomb ($1/r$) interaction.

The relevant length scale in a 2D HEG is its density parameter
$r_s=1/\sqrt{\pi n}$, where $n$ is the number density, while the
excitonic length scale is the exciton Bohr radius $a_0^* = 4\pi
  \epsilon_0 \epsilon \hbar^2/(\mu e^2)$, where $\mu=m_e
m_h/(m_e+m_h)$ is the reduced mass of the electron-hole pair, $m_e$
and $m_h$ are the electron and hole effective masses, and $\epsilon$ is
the static dielectric constant of the host material.  The energy scale
of excitonic systems is the exciton Rydberg (Ry$^*$), where $1\,{\rm Ry}^\ast
= \mu
e^4/(32\pi^2\epsilon_0^2\epsilon^2 \hbar^2)$.  We use Hartree atomic
units ($\hbar=|e|=m_e=4\pi\epsilon_0\epsilon=1$) unless otherwise
stated.

We use the \textsc{casino} code \cite{CASINO2} and VMC and DMC methods
to simulate systems containing 86 electrons and a single hole in a periodic
cell in the
presence of a uniform, neutralizing background charge density. In VMC,
expectation values are evaluated using a trial wave function
containing optimizable parameters.  The more accurate DMC method
projects out the lowest energy state with the same nodal surface as
the trial wave function \cite{FNA}.  The accuracy achieved for
expectation values of operators that commute with the Hamiltonian is
determined by the trial nodal surface, while the accuracy of other
expectation values and the statistical efficiency achieved are
influenced by the quality of the entire trial wave function.  Quantum
Monte Carlo (QMC) methods have previously been used to study related
systems including three-dimensional (3D) electron-hole gases
\cite{Littlewood}, 2D electron-hole gases \cite{Maezono_2013}, trions
in 2D materials \cite{Ganchev_2015,Mayers_2015}, excitons and
biexcitons in bilayer systems \cite{Tan_2005,Lee_2009,Bauer_2013}, and
positrons immersed in 3D electron gases
\cite{Boronski_2006,Drummond2}.

We use a trial wave function of
the form
\begin{equation}
  \Psi_{\rm T} ({\bf R}) =  e^{J({\bf R})}
  \Psi_{\rm S}\left[{\bf X}({\bf R})\right] \;,
\end{equation}
where ${\bf R}$ denotes the particle coordinates, $e^{J({\bf R})}$ is
a Jastrow factor that describes electron-electron and electron-hole
correlations \cite{Neil-J,Generic-J}, ${\bf X}({\bf R})$ is a set of
backflow-transformed coordinates \cite{Kwon,Inhom-BF}, and
\begin{equation}
\Psi_{\rm S}({\bf R}) =
  \det\left[\phi_i({\bf r}_j^\uparrow-{\bf r}_h)\right]
  \det\left[\phi_i({\bf r}_j^\downarrow-{\bf r}_h)\right] \;,
\end{equation}
is a product of Slater determinants containing orbitals that pair each
electron with the hole, where ${\bf r}_j^\sigma$ is the position
vector of the $j$th electron of spin $\sigma$ and ${\bf r}_h$ is the
position of the hole.
We use a novel form of flexible pairing orbital whose parameters are
optimized within VMC and which provides an accurate description of
electron-hole and electron-electron correlation,
\begin{equation}
\label{eq:orbital}
\phi_i({\bf r}) =
  \exp{[u_{G_i}(r)]}
  \exp\{i {\bf G}_i \cdot [r-\eta_{G_i}(r)] {\hat{\bf r}} \} \;,
\end{equation}
where ${\hat{\bf r}}$ is the unit vector in the direction of ${\bf
  r}$.  The orbital-dependent electron-hole Jastrow function $u_{G_i}$
and orbital-dependent electron-hole backflow function $\eta_{G_i}$,
where ${\bf G}_i$ is the $i$th shortest reciprocal lattice vector, are
described in the Supplemental Material \cite{supplemental}.  The free
parameters in the
backflow transformation, Jastrow factor, and orbitals are optimized
using energy minimization techniques \cite{Emin}.  We impose the
electron-electron and electron-hole Kato cusp conditions \cite{Kato}
via the Jastrow factor.

The pairing orbitals lower the VMC energies, roughly halving the
difference between the VMC and DMC relaxation energies, compared to
equivalent calculations with plane-wave orbitals.
The DMC energies are only lowered slightly by the use of pairing
orbitals; this insensitivity to changes in the nodal surface indicates
that the effect of the fixed-node approximation on the energies is
small.  A detailed comparison of results obtained using optimized
orbitals and plane-wave orbitals can be found in the Supplemental
Material \cite{supplemental}.

We use full shells of electrons in hexagonal simulation cells subject
to periodic boundary conditions.  In our production calculations we
use systems with $N_e=86$ electrons, which we deemed sufficiently
large after a finite-size-effect analysis involving systems of up to
$N_e=146$ electrons \cite{supplemental}.  We use a cell area of
$(N_e-1)\pi r_s^2$, so that the electron density far from the hole is
correct \cite{Bonev,Drummond1}.  We calculate the electron-hole
relaxation energy, also called the electron-hole correlation energy,
by subtracting the energy of a HEG of the same area containing the
same number of electrons.  The energy required to create an
electron-hole pair, for example via photoexcitation, is given by the
sum of the band gap, Fermi energy and relaxation energy, the latter
arising from the response of the electron gas to the point-particle
impurity.  Finite-size effects in the electron-hole relaxation energy
are small, and the pair-correlation functions (PCFs) are
well-converged with respect to system size \cite{supplemental}.  We
study systems with mass ratios $m_h/m_e=0.5$, 1, 2, 4, and 8.

\begin{figure}[tbh!]
  \begin{center}
    \includegraphics[clip,width=0.45\textwidth]{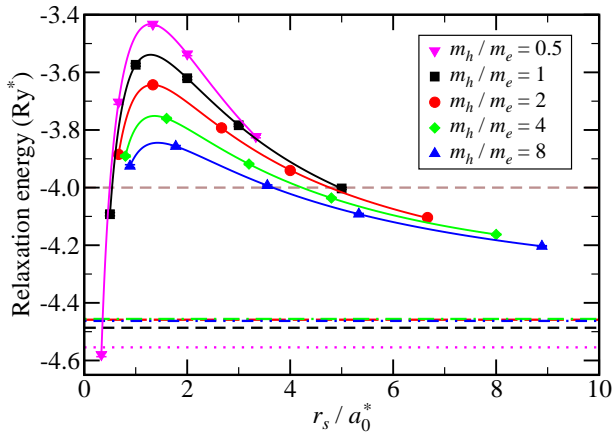}
    \caption{
      (Color online) Electron-hole relaxation energies for mass ratios
      $m_h/m_e=0.5$, 1, 2, 4, and 8, and HEG densities $r_s=1$, 2, 4,
      6, and 10 a.u.
      The brown dashed line indicates the isolated exciton energy at
      $-4~{\rm Ry}^*$, and the magenta dotted, black short-dashed, red
      long-dashed, green dot-dashed, and blue dot-dot-dashed lines show
      isolated trion energies at mass ratios $m_h/m_e=0.5$, 1, 2, 4, and 8,
      respectively.
      Fits of the electron-hole relaxation energies at each mass ratio
      have been constrained to tend to the respective trion energies at
      $r_s/a_0^*\to\infty$.
      Error bars are smaller than the symbols.
      \label{fig:relaxation-energy}
    }
  \end{center}
\end{figure}

The calculated electron-hole relaxation energies are shown in Fig.\
\ref{fig:relaxation-energy} together with the energy of the isolated
exciton, $E_X=-4$ Ry$^*$ \cite{Yang}.  We fit the electron-hole
relaxation energies for each mass ratio to functions that tend to the
energy of an isolated negative trion at $r_s/a_0^*\to\infty$, which is
the low carrier density limit of the electron-hole relaxation energy.
We evaluate isolated trion energies in separate DMC calculations;
numerical values are shown in Fig.\ \ref{fig:isolated-trion} and
tabulated in the Supplemental Material \cite{supplemental}.  Since
trions are composed of inequivalent particles, the wave function is
nodeless and DMC is exact in this case.  To obtain the $\mu/m_e\to0$
limit, we minimize the DMC energy as a function of the separation of
two fixed electrons, finding an equilibrium separation of
$0.51454(2)\,a_0^*$.  Our data are in good agreement with previous
results \cite{Zhu,Usukura} in the limiting cases studied in those
articles, and are compatible with recent calculations using the $1/r$
potential for trions in transition metal dichalcogenides
\cite{Mayers_2015}, although our energies are somewhat different to
those of Ref.\ \cite{Esser}.

At low densities the localized trion is similar to an electron,
and therefore the variation of the relaxation energy with $r_s$ is
dominated by the energy required to remove one electron from the HEG,
which is minus its Fermi energy.
At high densities the localized trion does not form, and at
$r_s/a_0^*\to 0$ the relaxation energy diverges towards $-\infty$,
as it does in three dimensions \cite{Arponen,supplemental}.

\begin{figure}[tbh!]
  \begin{center}
    \includegraphics[clip,width=0.45\textwidth]{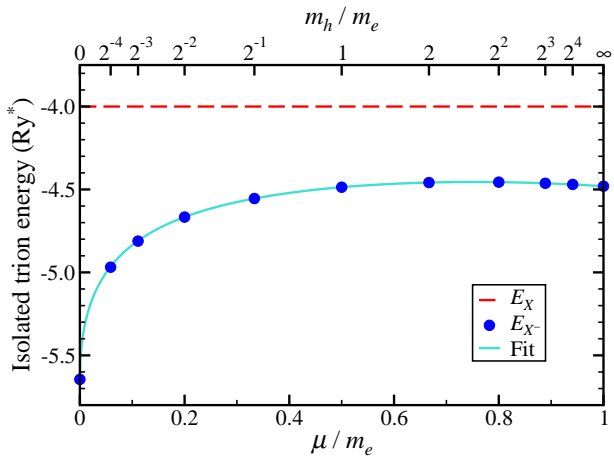}
    \caption{
      (Color online) Total energy of an isolated trion as a
      function of $\mu/m_e$ (circles).
      The energy of the neutral exciton is also shown (dashed line).
      Error bars are smaller than the symbols.
      \label{fig:isolated-trion}
    }
  \end{center}
\end{figure}

We calculate PCFs $g(r)$ using extrapolated estimation
\cite{Extrap,supplemental} to eliminate the leading-order dependence
on the trial wave function.  The VMC and DMC PCFs calculated using
plane-wave or optimized orbitals are almost identical, indicating the
robustness of the results.  Representative PCF data are shown in the
Supplemental Material \cite{supplemental}.  In Fig.\
\ref{fig:cumulative-PCF} we show integrated electron-hole PCFs which
give the total electron weight within a circle of radius $r$ centered
on the hole.  As the density is lowered, the $r^2$ behavior associated
with the electron gas is modified by the formation of a plateau at a
weight of two electrons, indicating the emergence of a trion.  Figure
\ref{fig:cumulative-PCF} shows that the trion radius decreases
somewhat for heavier holes; in excitonic units (not shown) these
curves coincide for $r \lesssim r_s$.  The electron-hole PCF tends to
that of an isolated trion at low density, while at high density the
free electrons screen the attractive potential from the hole,
preventing trion formation.  An intuitive explanation of the trends
seen in the relaxation energy is afforded by its close relationship
with the electron-hole PCF.  At low density, the relaxation energy
approaches the sum of the trion energy and the cost of the reduction
in density of the surrounding electron gas far from the trion caused
by the addition of the hole.  For $r_s \lesssim a_0^*$, the screening
charge becomes more localized close to the hole, increasing the
relaxation energy at high density \cite{supplemental}.

\begin{figure}[tbh!]
  \begin{center}
    \includegraphics[clip,width=0.45\textwidth]{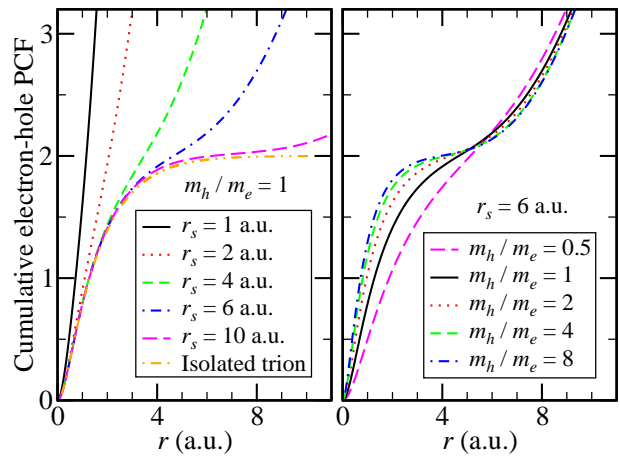}
    \caption{ (Color online) Cumulative (integrated) electron-hole
      PCFs normalized to show the total electron weight within a
      circle of radius $r$ centered on the hole.  Data are plotted for
      constant mass ratio $m_h/m_e=1$ and HEG densities $r_s=1$, 2, 4,
      6, and 10 a.u.\ (left panel), and for constant HEG density
      $r_s=6$ a.u.\ and mass ratios $m_h/m_e=0.5$, 1, 2, 4, and 8
      (right panel).
      \label{fig:cumulative-PCF}}
  \end{center}
\end{figure}

\begin{figure}[tbh!]
  \begin{center}
    \includegraphics[clip,width=0.45\textwidth]{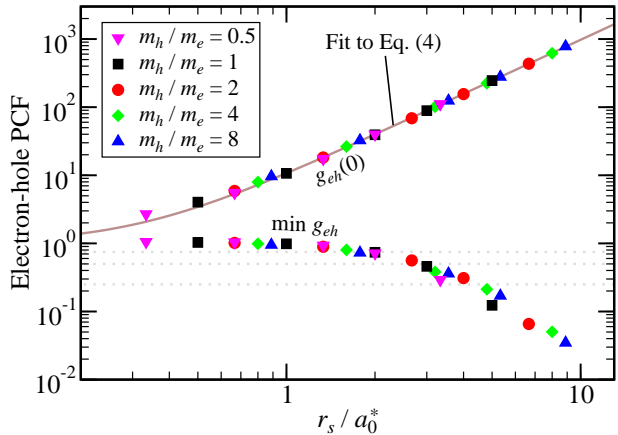}
    \caption{
      (Color online)
      Values of the electron-hole PCF at $r=0$ (on-top PCF) and at its
      first minimum.
      The solid line is a least-squares fit of the on-top PCF data to
      Eq.\ (\ref{eq:g_zero}).
      The PCF values 0.25, 0.5, and 0.75 are represented with dotted
      grey lines.
      Error bars are smaller than the symbols.
      Note the logarithmic scales.
      \label{fig:g-zero}}
  \end{center}
\end{figure}

The on-top electron-hole PCF $g_{eh}(0)$ is proportional to the rate of
electron-hole recombination.  It could also be used to create
semilocal two-component exchange-correlation functionals for use in
density functional theory calculations for modeling holes immersed in
inhomogeneous 2D systems \cite{g(0)-DFT}.  Considering the limits of
an exciton in a dilute electron gas and a trion without
electron-electron interaction in a dilute electron gas, we propose a
relation \cite{supplemental}:
\begin{equation}
\label{eq:g_zero}
g_{eh}(0) = c \mu^2 r_s^2 + 1 \;,
\end{equation}
where $c$ is a dimensionless parameter that is roughly
independent of $\mu$ and $r_s$ and takes values
slightly above the exciton limit of $c=8$.  We have extrapolated the
PCFs to $r=0$ and plotted the results against $\mu r_s$
($=r_s/a_0^*$) in Fig.\ \ref{fig:g-zero}.  Equation
(\ref{eq:g_zero}) fits the data well over the parameter space studied.
We obtain $c=9.742(7)$ from the fit; the variation
of $c$ with $\mu$ and $r_s$ is analyzed in
the Supplemental Material \cite{supplemental}.

The value of the electron-hole PCF at $r$ is the ratio of the
electronic density a distance $r$ from the hole to that of the
surrounding electron gas.
Thus, the value of the PCF at its first minimum, also shown in
Fig.\ \ref{fig:g-zero}, measures the degree of isolation
of the localized trion.
The minimum in the electron-hole PCF
develops at about $r_s\sim a_0^*$ and rapidly becomes
more pronounced as the carrier density decreases, with the
minimum electron density falling below 25\% of that of the surrounding
HEG by $r_s\sim 4.5 \, a_0^*$.

\begin{figure}[tbh!]
  \begin{center}
    \includegraphics[clip,width=0.45\textwidth]{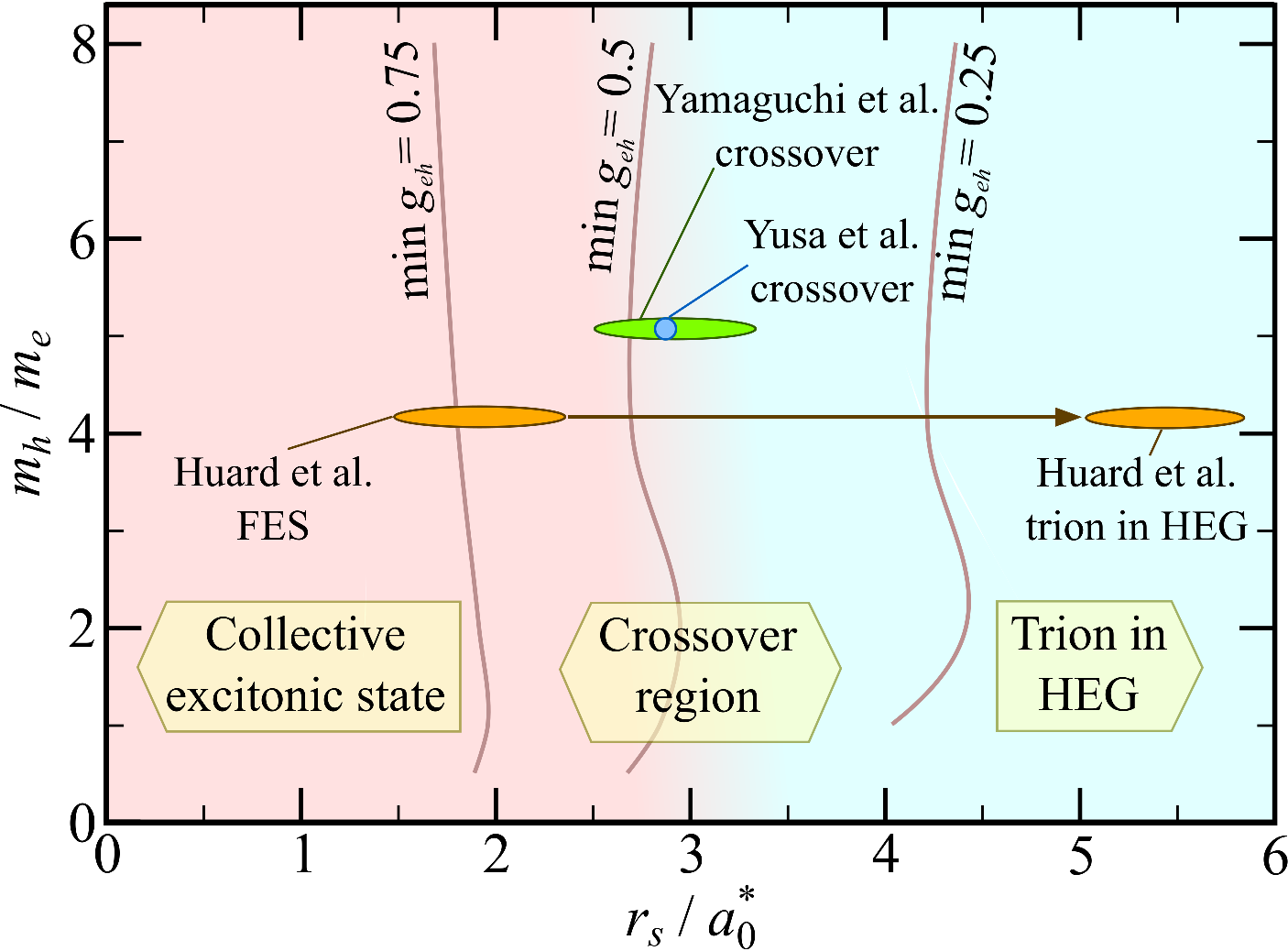}
    \caption{
      (Color online) Crossover of hole-in-HEG system from the high-density
      collective excitonic state to a localized trion immersed in a low
      density HEG as a function of
      mass ratio $m_h/m_e$ and density parameter $r_s/a_0^*$.
      Experimental data showing the evolution of absorption and
      photoluminescence spectra from the Fermi-edge singularity to discrete
      trion and exciton peaks are
      shown as colored areas, and are consistent with our results.
      \label{fig:phase-diagram}
    }
  \end{center}
\end{figure}

The gradual emergence of trions in our QMC calculations is compared to
experimental data from semiconductor quantum-well systems in Fig.\
\ref{fig:phase-diagram}.
This gradual crossover occurs in
a parameter range consistent with the absorption and photoluminescence
spectra seen experimentally.  The values of the dielectric constant
and effective masses used are given in the Supplemental Material,
together with an alternative picture of the crossover in the same
parameter range \cite{supplemental}.

\begin{figure}[tbh!]
  \begin{center}
    \includegraphics[clip,width=0.45\textwidth]{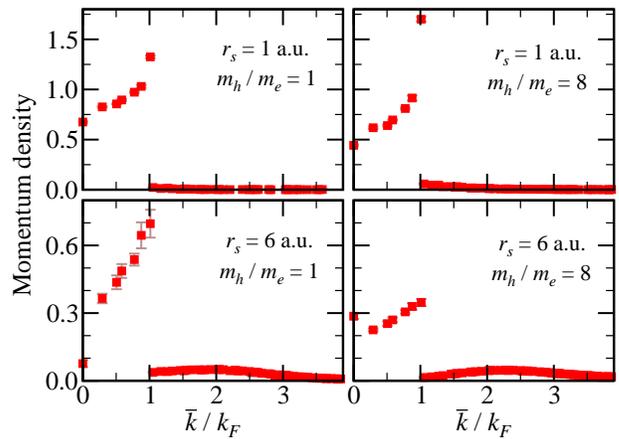}
    \caption{(Color online)
      Momentum density for four representative points of our ``phase
      diagram'': left column, $m_h/m_e=1$; right column, $m_h/m_e=8$;
      top row, $r_s=1$ a.u.; bottom row, $r_s=6$ a.u.
      Error bars are shown but are sometimes smaller than the symbols.
      Our results are normalized such that
      $\int_0^\infty 2\pi \bar{k} \rho(\bar{k}) \, d\bar{k} =\pi k_F^2$.
      \label{fig:rpmd}}
  \end{center}
\end{figure}

We have calculated the electron-hole center-of-mass momentum density
$\rho(\bar{k})$ by constraining one electron to remain on top of the
hole \cite{Drummond1,Drummond2,supplemental}.  Since it is not
possible to use extrapolated estimation for this quantity, we report
VMC results using the optimized orbitals.  The momentum density is
sensitive to the quality of the trial wave function, which has been
compensated for by using 500,000 configurations, a much larger number
than is required to converge the VMC energy, in order to minimize
noise during optimization.  In addition, each result is an average
over eight independently optimized wave functions \cite{supplemental}.
We have applied this method to compute the momentum densities of four
representative systems, as shown in Fig.\ \ref{fig:rpmd}.

At high density we obtain strong enhancement of momentum density just
below the Fermi edge, together with a small tail above the edge.  This
behavior was predicted theoretically by Carbotte and Kahana
\cite{Kahana, Carbotte} and recently demonstrated numerically by
Drummond \textit{et al.}\ \cite{Drummond2} for a positron in a 3D HEG
at metallic densities.  However, our results for 2D systems also show
the formation of a small, broad peak above the Fermi edge as the
density is lowered.  In contrast to the momentum density for
$\bar{k}<k_F$, this unusual peak is insensitive to small changes in
the wave function parameters and to the precise form of optimizable
wave function used.  The peak emerges gradually as the density is
lowered from that at which we estimate trion formation to begin, and
becomes higher and narrower, its center moving closer to the Fermi
edge, as the density is lowered to $r_s=10$ a.u.  We demonstrate that
this peak is associated with the formation of a trion in
the Supplemental Material \cite{supplemental}.

In conclusion, we have performed highly accurate QMC calculations for
a system containing a single hole immersed in a 2D electron gas.  Our
results demonstrate a crossover between a collective excitonic state
and a trion state as the density of the electron gas is lowered and as
the mass ratio is increased.  The electron-hole relaxation energy,
PCF, and electron-hole center-of-mass momentum density each show
evidence of the crossover.  The density and mass range in which trion
formation begins is in good agreement with recent experiments
\cite{Huard,Yusa,Yamaguchi}.

\begin{acknowledgments}
  The authors acknowledge financial support from the Engineering and
  Physical Sciences Research Council, U.K., under
  grant no.\ EP/J017639/1.
  Supporting research data may be freely accessed at
  \url{http://dx.doi.org/10.17863/CAM.565},
  in compliance with the applicable Open Access policies.
  Computational resources have been provided by the High Performance
  Computing Service of the University of Cambridge.
\end{acknowledgments}

\end{document}